\documentclass[showpacs,twocolumn,prl]{revtex4}
\usepackage{graphicx}
\usepackage{dcolumn}
\usepackage{bm}
\begin{document}
\title{
Origin and roles of a strong electron-phonon interaction\\
in cuprate oxide superconductors
}
\author{Fusayoshi J. Ohkawa}
\affiliation{Department of Physics, Faculty of
Science,  Hokkaido University, Sapporo 060-0810, Japan}
\email{fohkawa@phys.sci.hokudai.ac.jp}
\date{\today}
%
\begin{abstract} 
A strong electron-phonon interaction arises from the modulation
of the superexchange interaction by phonons.
It is responsible for the softening of
the  half breathing modes, 
as is studied in Phys.~Rev.~B {\bf 70}, 184514 (2004).
When antiferromagnetic spin fluctuations are developed
around  ${\bf Q}$, 
Cu-O bond stretching modes can also be soft around twice ${\bf Q}$
due to the electron-phonon interaction.
However, it can play no or only a minor role
in the binding of $d\gamma$-wave Cooper pairs.
\end{abstract}
\maketitle

It is an important issue to elucidate high critical temperature
(high-$T_c$) superconductivity in cuprate oxides 
discovered in 1986 \cite{bednorz}.
It occurs in the vicinity of the Mott metal-insulator transition or crossover.
In  1987,  a theory  is proposed  that
Gutzwiller's quasiparticles  
are bound into $d\gamma$-wave Cooper pairs
due to the superexchange interaction between nearest neighbors 
\cite{highTc1,highTc2}.
According to observed specific heat coefficients 
$\gamma \simeq 14~$mJ/K$^2$mol \cite{gamma1},
the bandwidth of quasiparticles is $W^*=0.3$-0.4~eV.
Since the superexchange interaction constant is as strong as
$J=-$(0.10-0.15)~eV \cite{SuperJ},
observed high $T_c$ can be easily reproduced.
High-$T_c$ superconductivity occurs in an intermediate-coupling regime
$|J|/W^* = 0.3\mbox{-}0.5$
for superconductivity, which is realized in the strong-coupling regime
for electron correlations.

It is also an important issue to elucidate the origin and roles
of a strong electron-phonon interaction, whose presence 
is implied by pieces of evidence:
the softening of the so called half breathing modes
around $\left(\pm \pi/a, 0\right)$ and $\left(0, \pm \pi/a\right)$
in the two-dimensional  (2D) Brillouin zone (BZ)
\cite{McQ1,Pint1,McQ2,Pint2,Braden},
with $a$ the lattice constant of CuO$_2$ planes,
the softening of Cu-O bond stretching modes
around $(\pm \pi/2a,0)$ and $(0,\pm \pi/2a)$ 
\cite{pintschovius,reznik},
kinks in the quasiparticle dispersion \cite{johnson,tsato}, and so on.
On the other hand,
observed small isotope shifts \cite{isotope} of 
$T_c$ imply that
the strong electron-phonon interaction plays only a minor role
in the occurrence of superconductivity. 

Since charge fluctuations are suppressed by strong electron correlations,
the conventional electron-phonon
interaction arising from charge-channel interactions must be
weak in cuprate oxides. 
An electron-phonon interaction arising  from spin-channel interactions 
can be strong when electron correlations are strong.
For example, it plays a significant role 
in the spin-Peierls effect. Then, 
a novel one is proposed in a previous paper \cite{novel-el-ph}:
one arising from the modulation of 
the superexchange interaction by phonons.
It can explain the softening of breathing modes
around $\left(\pm \pi/a, 0\right)$
and $\left(0, \pm \pi/a\right)$;
it is predicted that the softening must be small around 
$\left(\pm \pi/a, \pm \pi/a\right)$.
An attractive interaction arising from
the virtual exchange of a single phonon of the modes 
is very weak between nearest neighbors;
it is strong between next-nearest neighbors.
Then, it can play no significant role 
in the formation of $d\gamma$-wave Cooper pairs \cite{novel-el-ph}.
The observed small isotope shifts of $T_c$ can never contradict
the presence of this novel and strong electron-phonon interaction.
One of the purposes of this Letter is to explain the softening around
$\left(\pm \pi/2a, 0\right)$ and $\left(0, \pm \pi/2a\right)$
in 2D-BZ.

 
The superexchange interaction arises from the virtual exchange of pair 
excitations of $3d$ electrons across the upper Hubbard band (UHB)
and the lower Hubbard band (LHB) \cite{OhSupJ1}.
When their non-zero bandwidths are ignored,
the exchange interaction constant  
between nearest-neighbor Cu ions is given by
\begin{equation}\label{EqSuperJ}
J =-  \frac{4V^4}{(\epsilon_d-\epsilon_p+U)^2}
\left[\frac{1}{\epsilon_d-\epsilon_p+U}+\frac{1}{U} \right] ,
\end{equation}
with $V$ the hybridization matrix between nearest-neighbor $2p$ and
$3d$ orbits, $\epsilon_d$ and $\epsilon_p$ the depths of 
$3d$ and $2p$ levels, and $U$
the on-site repulsion between $3d$ electrons.
One may argue that parent cuprate oxides with no hole doping must be
charge-transfer insulators rather than Mott insulators
because doped holes reside mainly at O ions, so that
$\epsilon_p > \epsilon_d$. 
However, this argument disagrees with 
$\epsilon_{d}-\epsilon_{p} \simeq 1\mbox{~eV}$
predicted by band calculations \cite{band1,band2,band3}.
The preferential doping 
does not necessarily mean $\epsilon_p > \epsilon_d$, but it
simply means that the local charge susceptibility of $3d$ electrons is much
smaller than  that of $2p$ electrons. 
When we follow the band calculations and we use  
$V$=1.6~eV,  $\epsilon_{d}-\epsilon_{p}$=1~eV,  and  $U$=5~eV
%
\cite{OhSupJ1}, Eq.~(\ref{EqSuperJ}) gives
$J = - 0.27~\mbox{eV}$. 
This is about twice as large as the experimental 
$J = - (0.10\mbox{--}0.15)~\mbox{eV}$ \cite{SuperJ}.
This discrepancy is resolved when nonzero bandwidths of UHB and LHB
are considered \cite{OhSupJ1}.

Displacements of the $i$th Cu ion and the $[ij]$th O ion,
which lies between the $i$th and $j$th Cu ions, are given by
\begin{eqnarray}\label{EqDispCu}
{\bf u}_i \hspace{-3pt}&=&\hspace{-3pt}
 \sum_{\lambda{\bf q}}
 \frac{\hbar v_{d,\lambda{\bf q}} } 
{\sqrt{ 2NM_d \omega_{\lambda{\bf q}}} } 
e^{i{\bf q}\cdot{\bf R}_i}
{\bm \epsilon}_{\lambda{\bf q}} \left(
b_{\lambda-{\bf q}}^\dag \!+\! b_{\lambda{\bf q}} \right), 
\\  \label{EqDispO}
{\bf u}_{[ij]} \hspace{-3pt}&=&\hspace{-3pt}
\sum_{\lambda{\bf q}}
\frac{\hbar v_{p,\lambda{\bf q}}}
{\sqrt{2N M_p \omega_{\lambda{\bf q}}} } 
 e^{i{\bf q}\cdot {\bf R}_{[ij]} }
{\bm \epsilon}_{\lambda{\bf q}} \! \left(
b_{\lambda-{\bf q}}^\dag \!\!+\! b_{\lambda{\bf q}} \right), 
\end{eqnarray}
with  ${\bf R}_i$ and 
${\bf R}_{[ij]} \!=\! (1/2)({\bf R}_i \!+\! {\bf R}_j)$
positions of the $i$th Cu  and $[ij]$th O ions, 
$M_d$ and $M_p$ masses of Cu and O ions, 
$b_{\lambda{\bf q}}$ and 
$b_{\lambda-{\bf q}}^\dag$  
annihilation and creation operators of a phonon with 
a polarization $\lambda$ and a wave vector 
${\bf q}$ or $-{\bf q}$, $\omega_{\lambda{\bf q}}$ 
a phonon energy,  ${\bm \epsilon}_{\lambda{\bf q}}
=(\epsilon_{\lambda {\bf q},x},
\epsilon_{\lambda {\bf q},y},\epsilon_{\lambda {\bf q},z})$ 
a polarization
vector, and $N$ the number of unit cells. 
Here, we consider only longitudinal phonons; we assume 
$\epsilon_{\lambda{\bf q}} = (q_x,q_y,q_z)/|{\bf q}|$
for ${\bf q}$ within the first BZ.
The ${\bf q}$ dependence of
$v_{d,\lambda{\bf q}}$ and $v_{p,\lambda{\bf q}}$ is crucial.
For example, 
$v_{d,\lambda{\bf q}}=0$ and 
$v_{p,\lambda{\bf q}} = O(1)$   
for modes that bring no change in adjacent Cu-Cu distances.

Denoting creation and annihilation operators of $3d$ electrons 
at the $i$th site by 
$d_{i\sigma}^\dag$ and $d_{i\sigma}$
and those  of $3d$ electrons with wave number ${\bf k}$ by
$d_{{\bf k}\sigma}^\dag$ and $d_{{\bf k}\sigma}$,
we define {\it spin} operators by
${\bf S}_i $ $=$ $\sum_{\alpha\beta}   \frac1{2}
{\bm \sigma}^{\alpha\beta}  d_{i\alpha}^\dagger d_{i\beta}$ 
%
and 
${\bf S}({\bf q})$ $=$ $(1/\sqrt{N})$ $ \sum_{\bf k\alpha\beta} 
\frac1{2}{\bm \sigma}^{\alpha\beta} d_{({\bf k} +\frac{1}{2}{\bf q})
\alpha}^\dag   d_{({\bf k} -\frac{1}{2}{\bf q}) \beta}$, %
%
with  
$ {\bm \sigma} = \left(\sigma_x , \sigma_y ,\sigma_z \right)$
the Pauli matrixes.
Two types of electron-phonon interactions
arise from the modulations of $J$ by the vibrations of O and Cu ions:
\begin{eqnarray}\label{EqElPhP}
{\cal H}_p &=&
i C_p \sum_{\bf q} 
\frac{\hbar v_{p,\lambda{\bf q}}}
{\sqrt{2 N M_p \omega_{\lambda{\bf q}}}} 
\left(b_{\lambda-{\bf q}}^\dag + b_{\lambda{\bf q}} \right)
\nonumber \\ &&  \quad \times
\bar{\eta}_{s}({\bf q}) \sum_{\Gamma=s,d} 
\eta_{\Gamma}\left(\mbox{$\frac{1}{2}{\bf q}$}\right) 
{\cal P}_\Gamma({\bf q}) , 
\\ 
\label{EqElPhD}
{\cal H}_d &=&
i C_d \sum_{\bf q} 
\frac{\hbar v_{d,\lambda{\bf q}}}
{\sqrt{2 N M_d \omega_{\lambda{\bf q}}}} 
\left(b_{\lambda-{\bf q}}^\dag + b_{\lambda{\bf q}} \right)
\nonumber \\ &&  \quad \times
\sum_{\Gamma=s,d} 
\bar{\eta}_{\Gamma}({\bf q})
{\cal P}_\Gamma({\bf q}) , 
\end{eqnarray}
with $C_p$ and $C_d$ given in the previous paper
\cite{novel-el-ph}, 
%
%
%
%
\begin{eqnarray}
\bar{\eta}_{s}({\bf q}) \hspace*{-3pt} &=&\hspace*{-3pt}
2\left[ \epsilon_{\lambda {\bf q},x} \sin\left(q_x a/2\right) 
+  \epsilon_{\lambda {\bf q},y} \sin\left(q_y a/2\right)\right] ,
\quad \\
\bar{\eta}_{d}({\bf q})\hspace*{-3pt} &=& \hspace*{-3pt}
2\left[ \epsilon_{\lambda {\bf q},x} \sin\left(q_x a/2\right) 
-  \epsilon_{\lambda {\bf q},y} \sin\left(q_y a/2\right)\right] ,
\quad 
\\ && \hspace*{-0.5cm}
\eta_{s}({\bf k}) = \cos(k_xa) + \cos(k_ya),
\\ && \hspace*{-0.5cm}
\eta_{d}({\bf k}) = \cos(k_xa) - \cos(k_ya), 
\end{eqnarray}
being form factors, and
\begin{equation}\label{EqTwoSpin}
{\cal P}_\Gamma({\bf q}) = \frac1{2} 
\sum_{{\bf q}^\prime}\eta_{\Gamma}({\bf q}^\prime) \! \left[
 {\bf S}\left({\bf q}^\prime \!\!+\! \mbox{$\frac{1}{2}$}{\bf q}\right)
\!\cdot {\bf S}\left(-{\bf q}^\prime \!\!+\! \mbox{$\frac{1}{2}$}{\bf q}
\right)\right] ,
\end{equation}
being a {\it dual-spin} operator.
Here, the $x$ and $y$ axes are within CuO$_2$ planes, and
the $z$ axis is perpendicular to CuO$_2$ planes.
The $d$-$p$ model is approximately
mapped to the $t$-$J$ model \cite{ZhangRice}.
Then, we consider  the $t$-$J$-infinite $U$ model 
including ${\cal H}_p$ and ${\cal H}_d$ 
 on a square lattice:
\begin{eqnarray}\label{EqtJ}
{\cal H} &=& 
- \sum_{ij\sigma} t_{ij} d_{i\sigma}^\dag d_{j\sigma} 
-  \frac1{2}J 
\sum_{\left<ij\right>} ({\bf S}_i \cdot {\bf S}_j)  
\nonumber \\ && 
+  U_{\infty} \sum_{i}  
d_{i\uparrow}^\dag d_{i\uparrow}
d_{i\downarrow}^\dag d_{i\downarrow}
+ {\cal H}_p + {\cal H}_d, \quad 
\end{eqnarray}
with  $\left<ij\right>$ over nearest neighbors,
and $U_\infty$ an infinitely large on-site repulsion
to exclude double occupancy. 
Quasi-2D features are  considered in terms of
an anisotropy factor of spin fluctuations introduced below.

Every physical quantity is divided into  single-site and multi-site
ones. Calculating the single-site one is  reduced to
determining and solving selfconsistently the Anderson model,
which is an effective Hamiltonian for the Kondo problem. 
This is the single-site approximation (SSA)
that includes all the  single-site terms
 \cite{Mapping-1,Mapping-2,Mapping-3};
the SSA can also be formulated as
the dynamical-mean-field theory \cite{georges} or 
the dynamical-coherent-potential approximation \cite{kakehashi}.
Multi-site or intersite effects can be perturbatively considered
starting from a {\it unperturbed} state constructed in
the non-perturbative SSA theory.
Such a perturbative theory is nothing but a Kondo-lattice theory.

The irreducible polarization function in spin channels
is the sum of  single-site $\tilde{\pi}_s(i\omega_l)$,
which is the same as that of the Anderson model,
and multi-site  $\Delta\pi_s(i\omega_l,{\bf q})$:
%
$\pi_s(i\omega_l,{\bf q}) =
\tilde{\pi}_s(i\omega_l) +\Delta\pi_s(i\omega_l,{\bf q}) $.
%
Spin susceptibilities of the Anderson model and 
the $t$-$J$ model are given by
\begin{eqnarray}
\tilde{\chi}_s(i\omega_l) &=&
\frac{2\tilde{\pi}_s(i\omega_l)}{
1 - U_\infty \tilde{\pi}_s(i\omega_l)
},
\\
\chi_s(i\omega_l,{\bf q}) &=&
\frac{2\pi_s(i\omega_l,{\bf q})}{
1 - \left[\frac1{4}J({\bf q}) \!+\! U_\infty 
\right]\pi_s(i\omega_l,{\bf q})
},
\end{eqnarray}
with
%
$J({\bf q}) = 2 J \eta_s({\bf q}) $.
A physical picture for Kondo lattices is that
local spin fluctuations at different sites interact with each other
by an intersite exchange interaction.
Following this picture, we define 
an intersite exchange interaction 
$I_s(i\omega_l,{\bf q})$ by
\begin{equation}
\chi_s(i\omega_l,{\bf q}) =
\frac{\tilde{\chi}_s(i\omega_l)}{  
1 - \frac1{4}I_s(i\omega_l,{\bf q}) \tilde{\chi}_s(i\omega_l)} .
\end{equation}
Then, it follows that
%
$I_s(i\omega_l,{\bf q}) = J({\bf q})
+ 2 U_\infty^2 \Delta\pi_s(i\omega_l,{\bf q}) $.
%
The main term of 
$2 U_\infty^2 \Delta\pi_s(i\omega_l,{\bf q})$
is an exchange interaction arising from the exchange
of pair excitations of Gutzwiller's quasiparticles
\cite{FJO-disorder}.
It has a novel property that
its strength is proportional to the width of Gutzwiller's band.
It is antiferromagnetic (AF) 
when the chemical potential lies around the center of Gutzwiller's band
or the nesting of the Fermi surface is sharp enough;
it is ferromagnetic when the chemical potential lies around the top
or bottom of Gutzwiller's band. 

In cuprate oxides,
the exchange interaction $I_s(0,{\bf q})$  is AF 
and is maximal  around nesting wave numbers
of the Fermi surface. We assume that it is maximal at
${\bf Q}=\left(\pm 3\pi/4a, \pm\pi/a\right)$ and
$\left(\pm\pi/a, \pm 3\pi/4a \right)$
in 2D-BZ, and that the susceptibility is approximately given by
\begin{equation}\label{EqSus}
\chi_s\left(i\omega_l, {\bf Q} \!+\! {\bf q}\right)
= \frac{\chi_s(0,{\bf Q})\kappa^2}
{\displaystyle
\kappa^2 \!+\!(q_\parallel a)^2 \!+\! \delta^2 (q_z c)^2 \!+\! 
|\omega_l|/\Gamma_{AF}  },
\end{equation}
around each of ${\bf Q}$, where
%
${\bf q}_\parallel =(q_x,q_y)$, 
$\Gamma_{AF}$ is an energy scale of AF spin fluctuations, and
$c$ is the lattice constant along the $z$ axis.
The anisotropy factor $\delta$ is introduced to consider
quasi-2D AF spin fluctuations;
the correlation length within the $x$-$y$ plane is
$a/\kappa$ and that  along the $z$ axis is $\delta c /\kappa$.
A cut-off  $q_c=\pi/3a$ is introduced so that
$\chi_s\left(i\omega_l, {\bf Q} + {\bf q}\right)=0$
for $|q_x|>q_c$ or $|q_y|>q_c$.
The anisotropy of the lattice constants
plays no role when $\delta$
and $q_c$ are defined in these ways.

The Green function for
phonons  is given by
$D_\lambda(i\omega_l,{\bf q})=
2 \omega_{\lambda{\bf q}} / \bigl[
(i\omega_l)^2 \!-\! \omega_{\lambda{\bf q}}^2
\!+\! 2 \omega_{\lambda{\bf q}}
\Delta\omega_{\lambda{\bf q}}(i\omega_l) \bigr]$,
with $\Delta\omega_{\lambda{\bf q}}(i\omega_l)$
the renormalization or softening of phonons. 
%
Three types of fluctuations can be developed:
AF spin ones  due to $I_s(i\omega_l,{\bf q})$, and
$d\gamma$-wave superconducting (SC) and 
charge bond order (CBO) ones  due to 
a mutual interaction between quasiparticles, 
which is given by  
\begin{equation}\label{EqIdentity}
\chi_s(i\omega_l,{\bf q}) - \tilde{\chi}_s(i\omega_l) =
\frac{\frac1{4}I_s(i\omega_l,{\bf q})\tilde{\chi}_s^2(i\omega_l) }{
1 \!-\! \frac1{4}I_s(i\omega_l,{\bf q}) \tilde{\chi}_s(i\omega_l) 
},
\end{equation}
multiplied by $U_{\infty}^2$ and two single-site reducible vertex functions.
The softening around $2{\bf Q}$
occurs mainly due to AF spin ones around ${\bf Q}$;
$2{\bf Q}$'s are equivalent to
$(\pm \pi/2a,0)$ and $(0, \pm \pi/2a)$.
When only the AF spin ones are considered,
it follows that \cite{novel-el-ph}
\begin{equation}\label{EqSoftOmega}
\Delta\omega_{\lambda{\bf q}}(i\omega_l) = 
- \frac{\hbar^2}{2 M_p \omega_{\lambda{\bf q}} } 
\frac{3}{4^2}  \! \sum_{\Gamma\Gamma^\prime}
Y_{\Gamma}({\bf q}) Y_{\Gamma^\prime}({\bf q}) 
%
X_{\Gamma\Gamma^\prime}(i\omega_l,\!{\bf q}),
\end{equation}
with
%
%
\begin{eqnarray}&& \hspace*{-0.5cm}
Y_{\Gamma}({\bf q}) = \bar{\eta}_{s}({\bf q}) \hspace{-2pt}
\left[C_p v_{p,\lambda{\bf q}} 
\eta_{\Gamma} \left(\mbox{$\frac{1}{2}{\bf q}$}\right)   
\!+\!  C_d v_{d,\lambda{\bf q}}\sqrt{
\displaystyle \frac{M_p}{M_d} } \right], 
\\
\label{EqX} &&\hspace*{-0.5cm}
X_{\Gamma\Gamma^\prime}(i\omega_l,{\bf q})
=
\frac{k_B T}{N} \sum_{l^\prime{\bf p}}
\eta_{\Gamma}({\bf p}) \eta_{\Gamma^\prime}({\bf p}) 
%
\chi_s \! \left(i\omega_{l^\prime}, 
{\bf p} \!+\! \mbox{$\frac{1}{2}$}{\bf q}\right)
\nonumber \\ && \hspace*{2.0cm} \times
\chi_s \left(-i\omega_{l^\prime}-i\omega_l, 
-{\bf p} \!+\! \mbox{$\frac{1}{2}$}{\bf q}\right) .
\end{eqnarray}
In Eq.~(\ref{EqX}), two $\chi_s$'s appear because of
the {\it dual-spin} operator (\ref{EqTwoSpin}).
Since we are interested in Cu-O bond stretching modes around $2{\bf Q}$, 
we ignore vibrations of Cu ions; 
we assume that
$|C_d v_{d,\lambda{\bf q}}|\sqrt{M_p/M_d }=0$ and
\begin{equation}
|C_p v_{p,\lambda{\bf q}} | = 
c_p \mbox{~eV}/\mbox{\AA},
\end{equation}
where $c_p$ is a dimensionless constant and it is likely $c_p=O(1)$
\cite{novel-el-ph}.
Since the contribution from small ${\bf p}$ is large in 
Eq.~(\ref{EqX}), we consider only the contribution from 
the $\Gamma=s$ channel.
The softening is given by
\begin{equation}
\Delta\omega_{\lambda{\bf q}}(i\omega_l) =
- A_{\bf q} \Xi(i\omega_l,{\bf q}) ,
\end{equation}
with
\begin{eqnarray}\label{EqAq}
A_{\bf q} &=&
\frac{\hbar^2}{2 M_p \omega_{\lambda{\bf q}} } 
\frac{3}{4^2} \Gamma_{AF} \left[\chi_s(0,{\bf Q})\kappa^2\right]^2
|C_p v_{p,\lambda{\bf q}} |^2
\nonumber \\ &\simeq&
10 \times c_p^2 \frac{\Gamma_{AF}}{|t^*|} 
\left[\chi_s(0,{\bf Q}) \kappa^2 |t^*|
\right]^2 \mbox{~meV} ,
\end{eqnarray}
\begin{equation}\label{EqXi}
\Xi(i\omega_l, {\bf q}) =
\bar{\eta}_s^2({\bf q})
\eta_{s}^2 \left(\mbox{$\frac{1}{2}{\bf q}$}\right)   
\frac{X_{ss}(i\omega_l, {\bf q})}
{\Gamma _{AF} \left[\chi_s({\bf Q})\kappa^2\right]^2}.
\end{equation}
%
%
In Eq.~(\ref{EqAq}), $t^*$ is the
effective transfer integral between nearest neighbors for
quasiparticles, and
we assume $|t^*| \simeq W^*/8\simeq 40\mbox{-50 meV}$
and $\omega_{\lambda{\bf q}}=50\mbox{ meV}$.  
It is likely that
$\Gamma_{AF}/|t^*|= O(1)$ 
and $\chi_{s}(0,{\bf Q}) \kappa^2 |t^*| = O(1)$.
We note that $\Xi(i\omega_l, {\bf q})$ is defined as a dimensionless
quantity.
 
\begin{figure*}
\centerline{
\includegraphics[width=7.0cm]{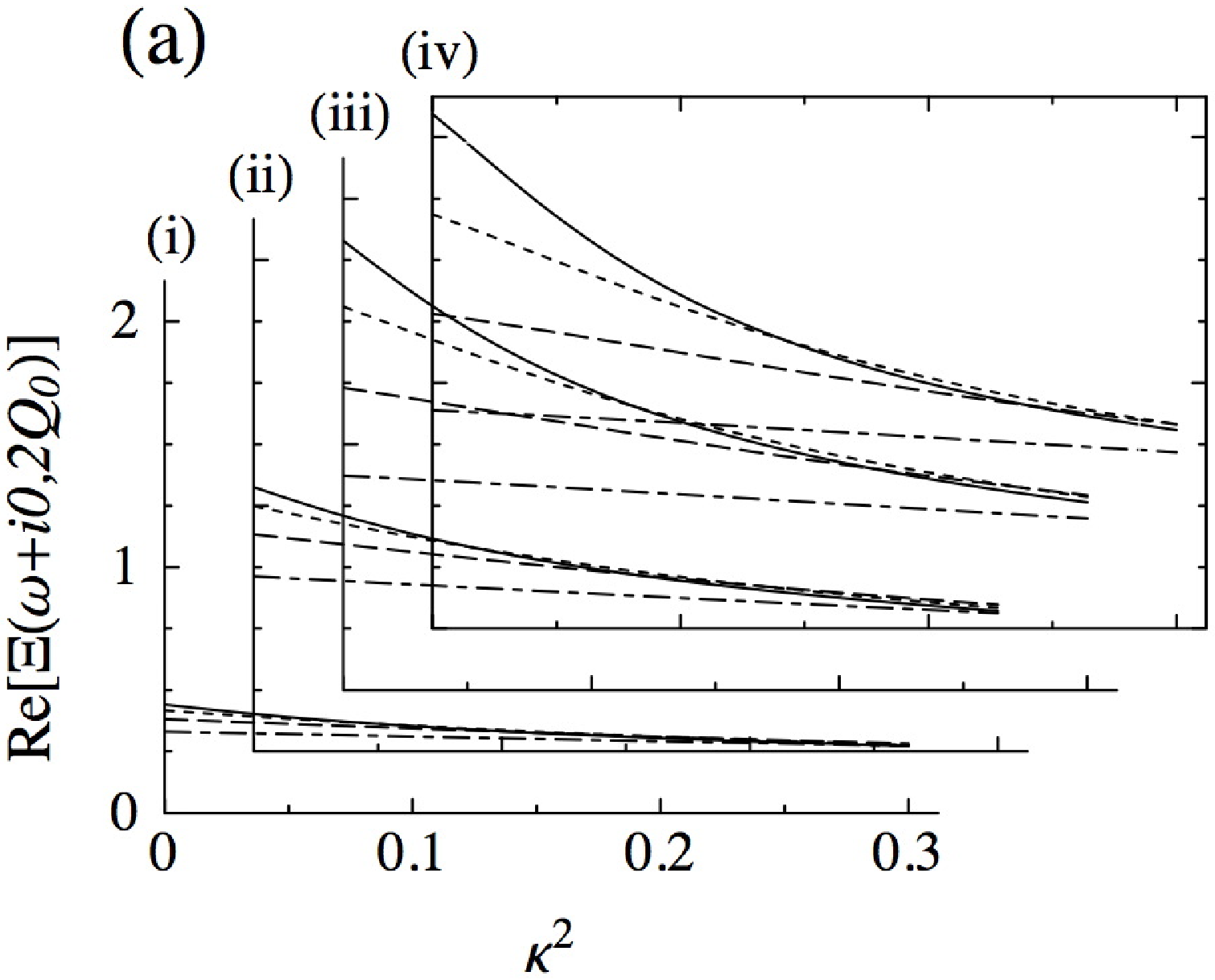}
\hspace{1.0cm}
\includegraphics[width=7.0cm]{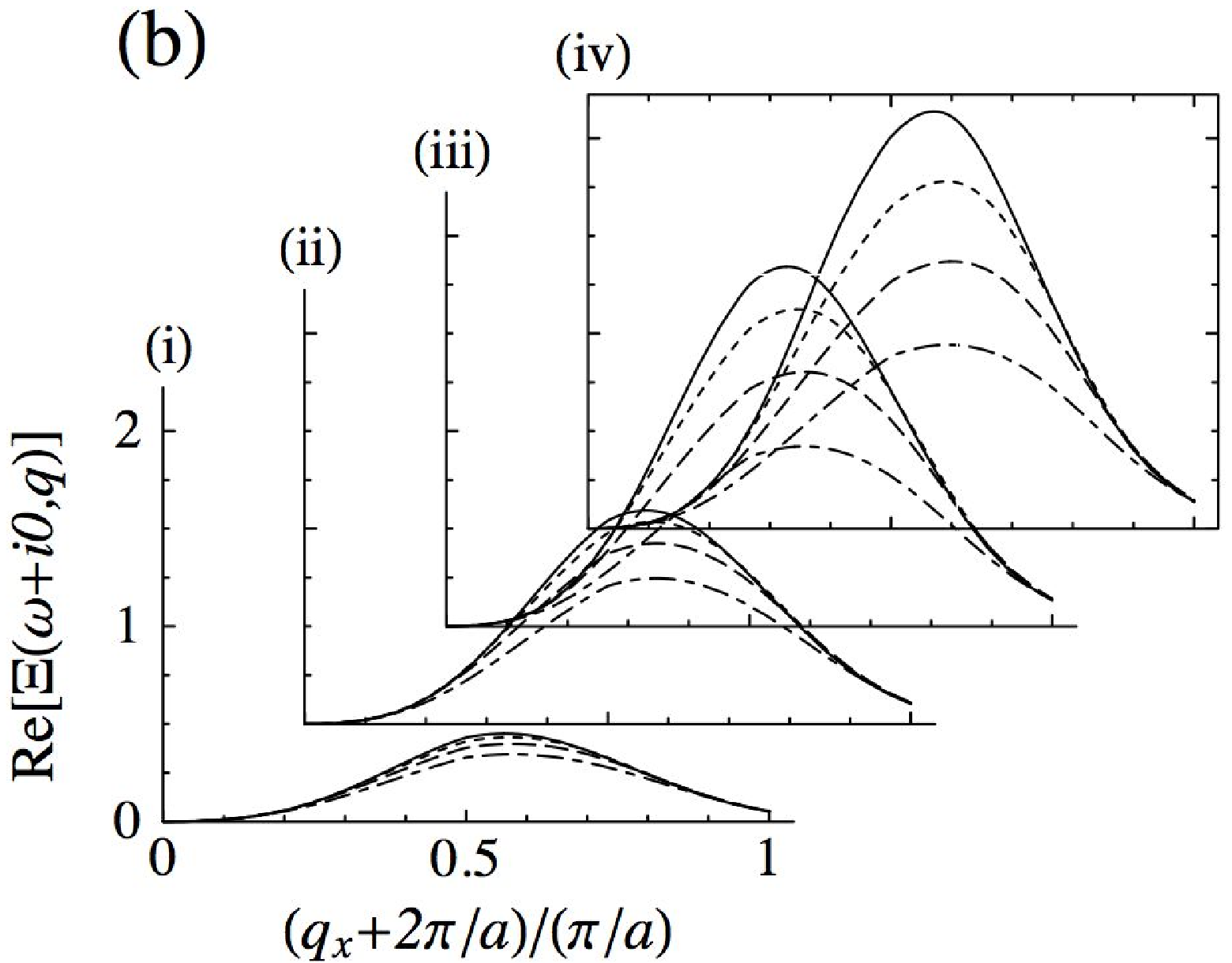}}
\caption[1]{
(a) $\mbox{Re}\bigl[\Xi(\omega+i0,2{\bf Q}_0)\bigr]$ 
as a function of $\kappa^2$, with ${\bf Q}_0=(-3\pi/4a,\pi/a)$,
and (b) $\mbox{Re}\bigl[\Xi (\omega+i0,{\bf q})\bigr]$ 
as a function of $q_x$ $(-2\pi/a \le q_x\le -\pi/a)$
for $q_y=2\pi/a$.  For the anisotropy factor, (i) $\delta=1$, 
(ii) $\delta=10^{-1/2}$, (iii) $\delta=10^{-1}$, and 
(iv) $\delta=10^{-3}$.
In each figure, solid, dotted, dashed, and dashed chain lines are for
$\omega/\Gamma_{AF}=$ 0.2, 0.4, 0.8, and 1.6, respectively. 
}
\label{fig_Xi}
\end{figure*}

We examine
the $\kappa$, $\delta$, and $\omega$
dependence of $\Xi(\omega+i0,{\bf q})$;  
$T=0$~K is assumed in the $\omega_{l^\prime}$ sum of Eq.~(\ref{EqX}),
and the softening around one of $2{\bf Q}$'s
or $2{\bf Q}_0$, with
${\bf Q}_0=(-3\pi/4a,\pi/a)$ in 2D-BZ, is considered;
$2{\bf Q}_0$ is equivalent to $(\pi/2a,0)$.
Figure~\ref{fig_Xi} shows 
$\mbox{Re}[\Xi(\omega+i0,{\bf q})]$
as a function of $\kappa^2$ or $q_x$
for several sets of $\delta$ and $\omega/\Gamma_{AF}$.
According to Fig.~\ref{fig_Xi},
$\mbox{Re}[\Delta\omega_{\lambda{\bf q}}
(\omega+i0) ]$
has a minimum around $2{\bf Q}_0$ as a 
function of ${\bf q}$; it is likely that its minimum value 
is as large as $-$(10-20) meV for
$\kappa^2 \ll 1$ and $\delta \ll 1$.

It is definite that
$\kappa^2 \ll1$ in the critical region, and
it is  certain
that $ \delta <10^{-3}$ for cuprate oxides.
Then, we propose that this second-harmonic effect of AF spin fluctuations
is mainly responsible for the observed softening \cite{pintschovius,reznik}
as large as $-$(10-20)~meV around $2{\bf Q}$.

Since the softening is small when $\kappa^2$ is large or
AF spin fluctuations are not developed, 
it must be small in the so called
over-doped cuprate oxides.
When AF spin fluctuations are developed similarly or differently between 
$\left(\pm 3\pi/4a, \pm\pi/a\right)$  and  
$\left(\pm\pi/a, \pm 3\pi/4a \right)$ 
because of the anisotropy of the Fermi surface within 2D-BZ,
the softening must also occurs similarly or differently between 
$\left(\pm\pi/2a,0\right)$ and $\left(0,\pm\pi/2a\right)$.
These two predictions are consistent with observations
\cite{pintschovius,reznik}.

The soft modes with $2{\bf Q}$ correspond to 
the so called stripes and checker-boards observed at low temperatures 
\cite{howald,Vershinin,Hanaguri,Momono,McElroy}.
Since a charge density wave (CDW) can appear 
following the complete softening, it should be examined if  
the stripes and checker-boards
can be explained in terms of the CDW.  
In general, the $2{\bf Q}$ component of 
the density of states, $\rho_{2{\bf Q}}(\varepsilon)$,  as a function of
$\varepsilon$ is composed of
symmetric and asymmetric ones with respect to 
the chemical potential or $\varepsilon=0$.
The asymmetric one is large
when CDW with $2{\bf Q}$ is  stabilized 
as a fundamental $2{\bf Q}$ effect.
According to an experiment \cite{howald},
the symmetric one is larger than the asymmetric one.
This  contradicts the scenario of CDW, even if 
the softening of the $2{\bf Q}$ modes is large and the $2{\bf Q}$
fluctuations are well developed; the amplitude of CDW  must be small,
even if it is stabilized. 
On the other hand, the symmetric one is large 
when the $2{\bf Q}$ modulation  is due to a second-harmonic effect
of a spin-density wave (SDW) with ${\bf Q}$ \cite{FJO-FFLO}. 
The observed almost symmetric 
$\rho_{2{\bf Q}}(\varepsilon)$ can be explained by
the second-harmonic effect of SDW. 
When stripes and checker-boards are really static orders, 
stripes must be due to single-${\bf Q}$ SDW and checker-boards
must be due to double-${\bf Q}$ SDW; 
magnetizations of the two waves must be orthogonal 
to each other in double-${\bf Q}$ SDW \cite{ortho1,ortho2}.

On the other hand, it is proposed  \cite{kiverson}
that a stripe or a checker-board at rather high temperatures 
must be an exotic ordered state, that is,
a fluctuating state in a quantum disordered phase.
It should be examined 
whether it is actually such an exotic state.
Another possibility is that it is 
a rather normal low-energy fluctuating state, whose
energy scale is as small as that of the soft phonons.
The other one is that it is
an disorder-induced SDW
\cite{FJO-disorder}; it must be
a rather simple but inhomogeneous SDW.

According to Eq.~(\ref{EqIdentity}),
two mechanisms of  attractive interactions,
the  spin-fluctuation one and  the exchange-interaction one,
are essentially the same as each other
in Kondo lattices.
However, the main part of the attractive interaction 
in cuprate oxides
is the superexchange interaction rather than an interaction
mediated by low-energy AF spin fluctuations.
Since it is as strong as $J=-(0.10$-0.15)~eV \cite{SuperJ},
observed high $T_c$ can be easily reproduced. 
Although the strong  electron-phonon interaction plays only a minor role 
in the formation of $d\gamma$-wave Cooper pairs,
a small isotope shift of $T_c$ can arise from the depression of
superconductivity by the $2{\bf Q}$ fluctuations.

In conclusion,
the strong electron-phonon interaction 
arising from the modulation of the superexchange interaction by phonons 
is strong in cuprate oxides superconductors.
It is responsible for the softening of
the half breathing modes
around $(\pm\pi/a, 0)$ and $(0,\pm\pi/a)$.
In the critical region of SDW, where
antiferromagnetic spin fluctuations are developed
around nesting wave numbers ${\bf Q}$ of the Fermi surface,
Cu-O bond stretching modes can also be soft
around $2{\bf Q}$.
The  softening is accompanied 
by the development of $2{\bf Q}$ or $2{\bf Q} \times 2{\bf Q}$
fluctuations, that is, stripe or checker-board fluctuations.
However, the observation that in ordered
stripe and checker-board states 
the $2{\bf Q}$ component of the density of states is almost symmetric 
with respect to the chemical potential 
can never be explained by CDW with $2{\bf Q}$ following
the complete softening of the $2{\bf Q}$ modes;
they can be explained by the second-harmonic effect 
of SDW with ${\bf Q}$.


\begin{thebibliography}{}
\bibitem{bednorz}
J. G. Bednortz and K. A. M\"{u}ller, 
Z. Phys. B {\bf 64}, 189 (1986).
%
\bibitem{highTc1} 
F. J. Ohkawa, 
Jpn. J. Appl. Phys. {\bf 26}, L652 (1987).
%
\bibitem{highTc2} 
F. J. Ohkawa, 
J. Phys. Soc. Jpn. {\bf 56}, 2267 (1987).
%
\bibitem{gamma1} 
J. W. Loram, et al.,
Phys. Rev. Lett. {\bf 71}, 1740 (1993).
%
%
%
%
%
%
%
\bibitem{SuperJ}
K. B. Lyons, et al.,
Phys. Rev. Lett. {\bf 60}, 732 (1988).
%
\bibitem{McQ1} 
R. J. McQueeney,  et al.,
Phys. Rev. Lett. {\bf 82}, 628 (1999).
%
\bibitem{Pint1}
L. Pintschovius and M. Braden,
Phys. Rev. B {\bf 60}, R15039 (1999).
%
\bibitem{McQ2}
R. J. McQueeney,  et al.,
Phys. Rev.  Lett. {\bf 87}, 077001 (2001).
%
\bibitem{Pint2}
L. Pintschovius,  et al.,
Phys. Rev. B {\bf 64}, 094510 (2001).
%
\bibitem{Braden} 
M. Braden, et al.,
Physica C {\bf 378}-{\bf 381}, 89 (2002).
%
\bibitem{pintschovius}
L. Pintschovius, et al.,
Phys. Rev. B {\bf 69}, 214506 (2004).
%
\bibitem{reznik} 
D. Reznik,  et al.,
Nature, {\bf 440}, 1170 (2006). 
%
\bibitem{johnson}
P. D. Johnson, et al.,
Phys. Rev. Lett {\bf 87}, 177007 (2001).
%
\bibitem{tsato}
T. Sato, et al.,
Phys. Rev. Lett. {\bf 91}, 157003 (2003).
%
\bibitem{isotope}
J. P. Franck, 
({\it Physical Properties of High Temperature Superconductors IV},
 ed. D.M. Ginsberg, World Scientific, Singapore, 1994)  p189. 
%
\bibitem{novel-el-ph}
F. J. Ohkawa,
Phys. Rev. B {\bf 70}, 184514 (2004).
%
%
\bibitem{OhSupJ1} 
F. J. Ohkawa, Phys. Rev. B {\bf 59}, 8930 (1999). 
%
\bibitem{band1}
T. Takegahara, H. Harima, and Y. Yanase,
Jpn. J. Appl. Phys. Part 1, {\bf 26}, L352 (1987)
%
\bibitem{band2}
N. Hamada,  et al.,
Phys. Rev. B {\bf 40}, 4442 (1989).
%
\bibitem{band3}
O. K. Andersen,  et al.,
Phys. Rev. B {\bf 49}, 4145 (1994).
%
\bibitem{ZhangRice}
F. C. Zhang and T. M. Rice,
Phys. Rev. B {\bf 37}, R3759 (1988). 
%
\bibitem{Mapping-1} 
F. J. Ohkawa, 
Phys. Rev. B {\bf 44}, 6812 (1991).
%
\bibitem{Mapping-2} 
F. J. Ohkawa, 
J. Phys. Soc. Jpn. {\bf 60}, 3218 (1991).
%
\bibitem{Mapping-3} 
F. J. Ohkawa, 
J. Phys. Soc. Jpn. {\bf 61}, 1615 (1992).
%
\bibitem{georges} 
A. Georges and G. Kotliar,
Phys. Rev. B {\bf 45}, 6479 (1992).
%
\bibitem{kakehashi}
Y. Kakehashi and P. Fulde,
Phys. Rev. B {\bf 69}, 45101 (2004).
%
%
%
%
\bibitem{FJO-disorder}
F. J. Ohkawa,
J. Phys. Soc. Jpn. {\bf 74}, 3340 (2005).
%
\bibitem{howald}
C. Howald,  et al.,
Phys. Rev. B {\bf 67}, 014533 (2003).
%
\bibitem{Vershinin} 
M. Vershinin, et al.,
Science {\bf 303}, 1995 (2004).
%
\bibitem{Hanaguri}
T. Hanaguri, et al.,
Nature {\bf 430}, 1001 (2004).
%
\bibitem{Momono}
N. Momono, et al.,
J. Phys. Soc. Jpn. {\bf 74}, 2400 (2005).
%
\bibitem{McElroy}
K. McElroy, et al.,
Phys. Rev. Lett. {\bf 94}, 197005 (2005).
%
\bibitem{FJO-FFLO}
F. J. Ohkawa,
Phys. Rev. B {\bf 73}, 092506 (2006).
%
\bibitem{ortho1}
F. J. Ohkawa,
J. Phys. Soc. Jpn. {\bf 67}, 535 (1998).
%
\bibitem{ortho2}
F. J. Ohkawa,
Phys. Rev. B {\bf 66}, 014408 (2002).
%
\bibitem{kiverson}
S. A. Kiverson, et al.,
Rev. Mod. Phys. {\bf 75}, 1202 (2003).
%
\end{thebibliography}
\end{document}